%% file: main.tex
\documentclass[11pt,leqno]{article}

\usepackage{times}
\usepackage{eepic,epic}
\usepackage{latexsym}
\usepackage{amsfonts}
\usepackage{amssymb}
\usepackage{amsmath}

\makeatletter%
\def\nottoobig#1{{\hbox{$\left#1\vcenter to1.111\ht\strutbox{}\right.\n@space$}}}
\makeatother%

\newlength{\filength}
\newsavebox{\gcbox}
\sbox{\gcbox}{\framebox[\filength]{\rule{0ex}{2ex}}}

\newcommand{\qedblob}{\mbox{\rule[-1.5pt]{5pt}{10.5pt}}}
\def\literalqed{{\ \nolinebreak\hfill\mbox{\qedblob\quad}}}

\def\qed{\literalqed}

\newtheorem{theorem}{Theorem}[section]

\newtheorem{definition}[theorem]{Definition}

\newtheorem{example}[theorem]{Example}

\newcommand\seq{\subseteq}

\newcommand{\pair}[1]{\mathopen\langle{#1}\mathclose\rangle}

\newcommand{\condition}{\,\nottoobig{|}\:}

\newcommand{\iso}[1]{\mbox{\rm{}ISO}({#1})}

\title{Computing Complete Graph Isomorphisms and Hamiltonian Cycles
from Partial Ones\thanks{This work was
supported in part by grant NSF-INT-9815095/DAAD-315-PPP-g\"{u}-ab.
The second author was supported in part by a Hei\-sen\-berg Fellowship
of the Deut\-sche For\-schungs\-ge\-mein\-schaft.
}
}

\author{
Andr\'{e} Gro\ss e\,\thanks{
Email: ${\tt grosse@informatik.uni\mbox{-}jena.de}$.
}
\\ Institut f\"ur Informatik
\\ Friedrich-Schiller-Universit\"at Jena
\\ 07740 Jena
\\ Germany
\and
J\"org Rothe\,\thanks{Corresponding author.  
Email: ${\tt rothe@cs.uni\mbox{-}duesseldorf.de}$.
}
\\ Mathematisches Institut
\\ Heinrich-Heine-Universit\"at D\"usseldorf
\\ 40225 D\"usseldorf
\\ Germany
\and
Gerd Wechsung\,\thanks{Email: 
${\tt wechsung@informatik.uni\mbox{-}jena.de}$.
}
\\ Institut f\"ur Informatik
\\ Friedrich-Schiller-Universit\"at Jena
\\ 07740 Jena
\\ Germany
} %

\date{June 18, 2001}

\begin{document}

\hbadness=3000%
\vbadness=10000 %

\bibliographystyle{alpha}

\maketitle

\let\BLS=\baselinestretch

\makeatletter
\newcommand{\niceonespacing}{\let\CS=\@currsize\renewcommand{\baselinestretch}{1.1}\tiny\CS}\newcommand{\nicetwospacing}{\let\CS=\@currsize\renewcommand{\baselinestretch}{1.2}\tiny\CS}
\newcommand{\nicethreespacing}{\let\CS=\@currsize\renewcommand{\baselinestretch}{1.3}\tiny\CS}
\newcommand{\singlespacingplusplus}{\let\CS=\@currsize\renewcommand{\baselinestretch}{1.35}\tiny\CS}
\newcommand{\nicefourspacing}{\let\CS=\@currsize\renewcommand{\baselinestretch}{1.4}\tiny\CS}
\newcommand{\nicefivespacing}{\let\CS=\@currsize\renewcommand{\baselinestretch}{1.5}\tiny\CS}
\newcommand{\nicesixspacing}{\let\CS=\@currsize\renewcommand{\baselinestretch}{1.6}\tiny\CS}
\makeatother
 
\pagestyle{plain}
\sloppy

\setcounter{footnote}{0}

\begin{abstract}
  We prove that computing a single pair of vertices that are mapped
  onto each other by an isomorphism $\phi$ between two isomorphic
  graphs is as hard as computing $\phi$ itself.  This result optimally
  improves upon a result of G\'{a}l et al.
  We establish a similar, albeit slightly weaker, result about computing
  complete Hamiltonian cycles of a graph from partial Hamiltonian cycles.

\vspace*{.5cm}
\noindent
\begin{tabular}{ll}
{\bf Key words:} & {\em computing from partial solutions; 
                   self-reducibility; graph isomorphisms;} \\
                 & {\em Hamiltonian cycles}
\end{tabular}
\end{abstract}

\clearpage

\section{Introduction}

Two of the most central and well-studied problems in NP are the graph
isomorphism problem and the Hamiltonian cycle problem.  The latter problem is
one of the standard NP-complete
problems~\cite{kar:b:reducibilities,gar-joh:b:int}.  In contrast, the graph
isomorphism problem currently is the most prominent candidate of a problem
that is neither in P nor NP-complete.  On the one hand, there is no efficient
algorithm known for solving this problem, despite a considerable effort in the
past to design such algorithms.  On the other hand, due to its well-known
lowness properties~\cite{sch:j:gi,koe-sch-tor:j:pplow}, the graph isomorphism
problem is very unlikely to be NP-complete.  For more information about the
graph isomorphism problem, we refer to the book by K{\"{o}}bler, Sch\"{o}ning,
and Tor\'{a}n~\cite{koe-sch-tor:b:graph-iso}.

Computational complexity theory and, in particular, the theory of
NP-completeness~\cite{gar-joh:b:int} traditionally is concerned with
the decision versions of problems.  For practical purposes, however,
to find or to construct a solution of a given NP problem is
much more important than merely to know whether or not a solution
exists.  For example, computing an isomorphism between two isomorphic
graphs (that is, solving the search version of the graph isomorphism
problem) is much more important for most applications than merely to know
that the graphs are isomorphic.  Therefore, much effort has been made
in the past to relate the complexity of solving the search problem to
the complexity of solving the corresponding decision problem.
This property is known as ``search reducing to decision,'' see, 
e.g.,~\cite{ehem-nai-ogi-sel:j:pselective} and the references cited
therein.  
The decisive property enabling search to reduce to decision for NP 
problems such as the graph isomorphism problem
is their self-reducibility.

The present paper builds on the recent work of
G\'{a}l, Halevi, Lipton, and 
Petrank~\cite{gal-hal-lip-pet:c:partial-solutions} who studied
a property that might be dubbed ``complete search reducing
to partial search.''~~They showed for various NP problems $A$ that,
given an input $x \in A$, computing a small fraction of a solution for
$x$ is no easier than computing a complete solution for~$x$.  For
example, given two isomorphic graphs, computing roughly
logarithmically many pairs of vertices that are mapped onto each other
by a complete isomorphism $\phi$ between the graphs is as hard as
computing $\phi$ itself.  

As G\'{a}l et al. note, their results have
two possible interpretations.  Positively speaking, their results say that to
efficiently solve the complete search problem it is enough to come up
with an efficient algorithm for computing only a small part of a solution.
Negatively speaking, their results say that constructing even a small part of
a solution to instances of hard problems
also appears to be a very difficult task.  The work of
G\'{a}l et al.~\cite{gal-hal-lip-pet:c:partial-solutions} also has
consequences with regard to fault-tolerant computing (in particular,
for recovering the complete problem solution when parts of it are lost
during transmission), and for constructing robust proofs of membership.

The present paper makes the following contributions.  Firstly, we
improve the above-mentioned result of G\'{a}l et
al.~\cite{gal-hal-lip-pet:c:partial-solutions} by showing that
computing even a single pair of vertices that are mapped onto each other by
a complete isomorphism $\phi$ between two isomorphic graphs is as hard
as computing $\phi$ itself.  This result is a considerable
strengthening of the previous result and an optimal improvement.
Interestingly, the self-reducibility of the graph isomorphism problem
is the key property that makes our stronger result possible.

Secondly, we obtain a similar, albeit somewhat weaker, result about
computing complete Hamiltonian cycles of a given graph from accessing to
partial information about the graph's Hamiltonian cycles.

\section{Computing Complete Graph Isomorphisms from Partial Ones}

G\'{a}l et al.~\cite{gal-hal-lip-pet:c:partial-solutions} prove the
following result.  Suppose there exists a function oracle $f$ that,
given any two isomorphic graphs with $m$ vertices each, outputs a part
of an isomorphism between the graphs consisting of at least $(3 +
\epsilon)\log m$ vertices for some constant $\epsilon > 0$.  Then,
using the oracle~$f$, one can compute a complete isomorphism between any
two isomorphic graphs in polynomial time.  

We improve their
result by showing the same consequence under the weakest assumption
possible: Assuming that we are given a function oracle that provides
{\em only one vertex pair\/} belonging to an isomorphism between two
given isomorphic graphs, one can use this oracle to
compute complete isomorphisms between two
isomorphic graphs in polynomial time.
Thus, our improvement of the previous result by G\'{a}l 
et al.~\cite{gal-hal-lip-pet:c:partial-solutions} is optimal.

\begin{definition}
  For any graph~$G$, the vertex set of $G$ is denoted by~$V(G)$, and the edge
  set of $G$ is denoted by~$E(G)$.
  
  Let $G$ and $H$ be undirected and simple graphs, i.e., graphs with no
  reflexive and multiple edges.
  
  An {\em isomorphism\/} between $G$ and $H$ is a bijective mapping $\phi$
  from $V(G)$ onto $V(H)$ such that, for all $x,y \in V(G)$,
\[
\{x,y\} \in E(G) \,\Longleftrightarrow\, \{\phi(x),\phi(y)\} \in E(H).
\]  
Let $\iso{G,H}$ denote the set of isomorphisms between $G$ and~$H$.
\end{definition}

We now state our main result.

\begin{theorem}
\label{thm:graphiso}
  Suppose there exists a function oracle $f$ that, given any two
  isomorphic graphs $\hat{G}$ and~$\hat{H}$, outputs two vertices $x
  \in V(\hat{G})$ and $y \in V(\hat{H})$ with $\hat{\phi}(x) = y$, for
  some isomorphism $\hat{\phi}$ from $\iso{\hat{G},\hat{H}}$.  

  Then, there is a recursive procedure $g$ that, given any two
  isomorphic graphs $G$ and~$H$, uses the oracle $f$ to construct a complete
  isomorphism $\phi \in \iso{G,H}$ in polynomial time.
\end{theorem}

Before proving Theorem~\ref{thm:graphiso}, we
explain the main
difference between our proof and the proof of G\'{a}l et
al.~\cite{gal-hal-lip-pet:c:partial-solutions}.  Crucially, to make
their recursive procedure terminate, they ensure in their
construction that the (pairs of)
graphs they construct are of strictly decreasing size 
in each loop of the procedure.
In contrast, for our algorithm this strong requirement is not 
necessary to make the procedure terminate.

Let us informally explain why.
Our algorithm is inspired by
the known self-reducibility algorithm for the graph isomorphism
problem; see, e.g.,~\cite{koe-sch-tor:b:graph-iso}.
The notion of self-reducibility has been thoroughly studied by 
many authors; we refer the reader to the work of 
Schnorr~\cite{sch:j:self-reducibility,sch:c:self-transformable},
Meyer and Paterson~\cite{mey-pat:t:int},
Selman~\cite{sel:j:natural}, and 
Ko~\cite{ko:j:self-reducibility}, and 
to the excellent survey by Joseph and
Young~\cite{jos-you:b:internal-structure} 
for an overview and for pointers to the literature.

Informally speaking, 
a self-reduction for a problem $A$ is a computational procedure for
solving~$A$, where the set $A$ itself may be accessed as an oracle.
To prevent this notion from being trivialized, one requires that $A$
cannot be queried about the given input itself; usually, only queries
about strings that are ``smaller'' than the input string are allowed.
When formally defining what precisely is meant by ``smaller,''
most self-reducibility
notions---including those studied by the above-mentioned 
researchers---employ the useful concepts of
``polynomially well-founded'' and ``length-bounded''
partial orders, rather than being based simply on the lengths of strings.
This approach is useful in order
to ``obtain full generality and to preserve the concept under
polynomially computable
isomorphisms''~\cite[p.~84]{jos-you:b:internal-structure}, see
also~\cite{mey-pat:t:int,sel:j:natural}.    
That means that the strings queried in a self-reduction may be 
{\em larger in length\/} than the input strings as long as they are 
{\em predecessors in a polynomially well-founded and 
length-bounded partial order}.
It is this key property that makes our algorithm terminate without
having to ensure in the
construction that the (pairs of)
graphs constructed are of strictly decreasing size 
in each loop.

Here is an intuitive description of how our algorithm works.  Let $G$ and $H$
be the given isomorphic graphs.  The function oracle will be invoked in each
loop of the procedure to yield any one pair of vertices that are mapped onto
each other by some isomorphism between the graphs as yet constructed.  
However, if we
were simply deleting this vertex pair, we would obtain new graphs $\hat{G}$
and $\hat{H}$ such that $\iso{\hat{G},\hat{H}}$ might contain some isomorphism
not compatible with $\iso{G,H}$, which means it cannot be extended to an
isomorphism in $\iso{G,H}$.  That is why our algorithm will attach cliques of
appropriate sizes to each vertex to be deleted, and the deletion of this 
vertex, and of the clique attached to it, will be delayed until some 
subsequent loop of the procedure.  That is, the (pairs of)
graphs we construct may increase in size in some of the loops, and yet the
procedure is guaranteed to terminate in polynomial time.

We now turn to the formal proof.

\medskip

\noindent
{\bf Proof of Theorem~\ref{thm:graphiso}.}~~Let 
$G$ and $H$ be two given isomorphic graphs with $n$
vertices each.  Let $f$ be a function oracle as in the theorem.  We describe
the recursive procedure $g$ that computes an isomorphism $\phi \in
\iso{G,H}$.  Below, we use variables $\hat{G}$ and $\hat{H}$ to
denote (encodings of) graphs obtained from $G$ and $H$ according
to~$g$, and we refer to the vertices of $G$ and $H$ as the {\em old\/}
vertices and to the vertices of $\hat{G} - G$ and $\hat{H} - H$ as the {\em
new\/} vertices.

On input $\pair{G,H}$, the algorithm $g$ executes the following steps:
\begin{enumerate}
\item Let $\hat{G} = G$ and $\hat{H} = H$, and set $i$ to $n =
  ||V(G)||$.  Let $\phi \seq V(G) \times V(H)$ be a set variable that,
  eventually, gives
  the isomorphism between $G$ and $H$ to be constructed.  Initially,
  set $\phi$ to the empty set.
  
\item Query $f$ about the pair $(\hat{G},\hat{H})$.
  Let $(x,y)$ be the vertex pair returned by $f(\hat{G},\hat{H})$, 
  where $x \in V(\hat{G})$ and $y \in V(\hat{H})$ and $\hat{\phi}(x) = y$ 
  for some isomorphism $\hat{\phi} \in \iso{\hat{G},\hat{H}}$.
\label{step:iterate}

\item Consider the following two cases:
\label{step:cases}
\begin{description}
\item[Case~\ref{step:cases}.1:] $x \in V(G)$ is an old vertex.  

We distinguish the following two cases:
\begin{enumerate}
\item
  $y$ is also an old vertex (in~$H$).
  
  Set $\phi$ to $\phi \cup \{(x,y)\}$.  Modify the graphs $\hat{G}$ and
  $\hat{H}$ as follows.

  Delete~$x$, all new neighbors of~$x$, and all edges incident to either
  of these vertices from~$\hat{G}$.  Attach to each old neighbor $x'
  \in V(G)$ of $x$ a copy of a clique $C_{i,x'}$ consisting of $i-1$
  new vertices each of which is connected with $x'$ by an edge; hence,
  the graph induced by $V(C_{i,x'}) \cup \{x'\}$ forms an $i$-clique.
  Make sure that all the new clique vertices are pairwise disjoint and disjoint
  with (the old) graph~$\hat{G}$.  Call the resulting graph (the
  new)~$\hat{G}$.

  Modify $\hat{H}$ in the same way: Delete $y$ and all
  new neighbors of $y$ from~$\hat{H}$, and extend each old neighbor
  $y' \in V(H)$ of $y$ to a clique consisting of the $i$ vertices
  $V(C_{i,y'}) \cup \{y'\}$.
  
\item
  $y$ is a new vertex in~$H$.
  
  Let $\tilde{y} \in V(H)$ be the unique old vertex adjacent to~$y$, i.e.,
  $y$ is a member of the clique $C_{j,\tilde{y}}$ that was previously 
  attached to $\tilde{y}$ in the $(j-n+1)$th loop, where $n \leq j < i$.  
  Note that the size of the clique $C_{j,\tilde{y}} \cup \{\tilde{y}\}$ 
  equals~$j$.
  Since $\hat{\phi}(x) = y$, the old vertex $x$ must belong to the clique 
  $C_{j,x} \cup \{x\}$ of size $j$ and, thus, cannot have any old neighbors 
  in~$\hat{G}$.  It follows that $\tilde{y}$ is also not adjacent to 
  any old vertex in the current graph~$\hat{H}$.  That is, both the clique
  $C_{j,x} \cup \{x\}$ and the clique $C_{j,\tilde{y}} \cup \{\tilde{y}\}$
  are connecting components of their graphs~$\hat{G}$ and $\hat{H}$,
  respectively.  

  Set $\phi$ to $\phi \cup \{(x,\tilde{y})\}$.  Modify the graphs $\hat{G}$
  and $\hat{H}$ by deleting the cliques
  $C_{j,x} \cup \{x\}$ and $C_{j,\tilde{y}} \cup \{\tilde{y}\}$.
\end{enumerate}
  
Set $i$ to $i+1$.
  
\item[Case~\ref{step:cases}.2:] $x \not\in V(G)$ is a new vertex
  in~$\hat{G}$.  

  It follows that $x$ is a member of a clique $C_{j,\tilde{x}}$,
  where $n \leq j < i$, that was previously attached to some old vertex
  $\tilde{x} \in V(G)$ in the $(j-n+1)$th loop.  Also, by
  construction, $\tilde{x}$ is the only old vertex adjacent to~$x$.
  Similarly, it holds that $y$ is a member of a clique $C_{j,\tilde{y}}
  \cup \{\tilde{y}\}$ in $\hat{H}$ with a uniquely determined old 
  vertex $\tilde{y} \in V(H)$.

  If $y = \tilde{y}$, then this case reduces to 
  Case~\ref{step:cases}.1(a), with $x$ being replaced by $\tilde{x}$.

  If $y \neq \tilde{y}$, then
  $\hat{\phi}(x) = y$ implies that $\hat{\phi}(\tilde{x}) = \tilde{y}$
  and, thus, that $\tilde{x}$ and $\tilde{y}$ have the same number of
  old neighbors.
  Hence, this case also reduces to Case~\ref{step:cases}.1(a), with $x$ 
  being replaced by $\tilde{x}$ and $y$ being replaced by~$\tilde{y}$.
\end{description}

\item If there are no vertices left in $\hat{G}$ and~$\hat{H}$,
  output~$\phi$, which gives a complete isomorphism between $G$
  and~$H$.  Otherwise, go to Step~\ref{step:iterate}.
\end{enumerate}

As alluded to in the above informal description of the algorithm,
the intuition behind introducing cliques of increasing sizes in the
construction is to keep the isomorphisms $\hat{\phi} \in
\iso{\hat{G},\hat{H}}$ compatible with $\phi \in \iso{G,H}$ when
vertices from $G$ and $H$ are deleted.  That is, we want to preclude
the case that deleting $x \in V(G)$ and $y \in V(H)$ results in
reduced graphs $\hat{G}$ and $\hat{H}$ such that there is some
$\hat{\phi} \in \iso{\hat{G},\hat{H}}$---and our oracle $f$ might pick
some vertex pair corresponding to such a~$\hat{\phi}$---that cannot be
extended to $\phi \in \iso{G,H}$.

The following example illustrates this intuition and shows how the algorithm
works.

\begin{example}
\label{ex}
Figure~\ref{fig:example}
gives an example of a pair
of isomorphic graphs $G$ and $H$ with $\iso{G,H} = \{\phi_1 , \phi_2\}$, 
where 
\begin{eqnarray*}
\phi_1 = \left( 
\begin{array}{ccccc}
1 & 2 & 3 & 4 & 5 \\
1 & 5 & 4 & 3 & 2 
\end{array}
\right)
& \mbox{ and } & 
\phi_2 = \left( 
\begin{array}{ccccc}
1 & 2 & 3 & 4 & 5 \\
5 & 1 & 4 & 3 & 2
\end{array}
\right).
\end{eqnarray*}

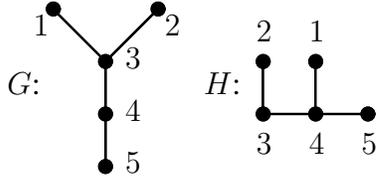
\begin{figure}[p]
\setlength{\unitlength}{1cm}
\begin{minipage}[t]{5.8cm}
\begin{picture}(5.8,3.2)
\input{example.eepic}
\end{picture}\par
\caption{Two graphs $G$ and $H$ with $\iso{G,H} = \{\phi_1 , \phi_2\}$.
\label{fig:example}
}
\end{minipage}
\hfill
\begin{minipage}[t]{5.8cm}
\begin{picture}(5.8,3.2)
\input{badprocedure.eepic}
\end{picture}\par
\vspace*{1.8ex}
Figure~~2:~~Two graphs $\widehat{G}$ and
$\widehat{H}$ for which $\iso{\widehat{G},\widehat{H}}$ contains
isomorphisms not compatible with the pair $(5,2)$.
\end{minipage}
\end{figure}
\setcounter{figure}{2}

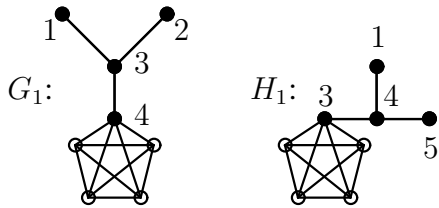
\begin{figure}[htp]
\setlength{\unitlength}{1cm}
\begin{minipage}[t]{5.8cm}
\begin{picture}(5.8,3.2)
\input{goodprocedure.eepic}
\end{picture}\par
\caption{Two graphs $G_1$ and $H_1$ obtained from $G$ and $H$ 
  according to $g$ when $f(G,H)$ returns $(5,2)$.
\label{fig:three}
}
\end{minipage}
\hfill
\begin{minipage}[t]{5.8cm}
\begin{picture}(5.8,3.2)
\input{goodprocedure2.eepic}
\end{picture}\par
\vspace*{1.8ex}
Figure~~4:~~Two graphs $G_2$ and $H_2$ that result from 
$f(G_1,H_1) = (1,5)$.
\end{minipage}
\end{figure}

\setcounter{figure}{4}
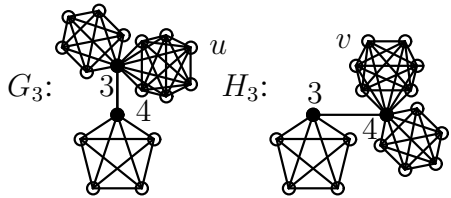
\begin{figure}[p]
\setlength{\unitlength}{1cm}
\begin{minipage}[t]{5.8cm}
\begin{picture}(5.8,3.2)
\input{goodprocedure3.eepic}
\end{picture}\par
\caption{Two graphs $G_3$ and $H_3$ that result from 
$f(G_2,H_2) = (2,1)$.}
\end{minipage}
\hfill
\begin{minipage}[t]{5.8cm}
\begin{picture}(5.8,3.2)
\input{goodprocedure4.eepic}
\end{picture}\par
\vspace*{1.8ex}
Figure~~6:~~Two graphs $G_4$ and $H_4$ that result from  
  $f(G_3,H_3) = (u,v)$.
\end{minipage}
\end{figure}

Suppose that the function oracle~$f$, when queried about the pair~$(G,H)$,
returns, e.g., the vertex pair~$(5,2)$.
If we were simply deleting the vertex~$5$
from $G$ and the vertex~$2$ from~$H$, then we would obtain graphs 
$\widehat{G}$ and $\widehat{H}$ such that $\iso{\widehat{G},\widehat{H}}$
contains six isomorphisms only two of which are compatible with the
pair~$(5,2)$; see Figure~2.  
But then~$f$, when queried about the pair $(\widehat{G},\widehat{H})$, 
might pick, e.g.,
the vertex pair $(4,5)$, which belongs neither to $\phi_1$ nor to~$\phi_2$.

To preclude cases like this, our algorithm attaches cliques of size~5 to
the vertex~$4$ in $G$ and to the vertex~$3$ in~$H$; 
see Figure~\ref{fig:three}.
Old vertices are represented by full circles and new vertices by empty
circles. 
Note that each $\phi \in \iso{G_1,H_1}$ is compatible with the vertex
pair $(5,2)$ from $\phi_1 , \phi_2 \in \iso{G,H}$. 

Figure~\ref{fig:three} through Figure~6 show
how~$g$, on input $\pair{G,H}$, continues to work for a 
specific sequence of oracle answers from~$f$.
In Figure~6, the only old vertex left in $G_4$
is the vertex~$4$, and the only old vertex left in $H_4$ is the vertex~$3$.
Hence, whichever vertex pair~$f$ when queried
about $(G_4,H_4)$ picks, $g$ maps the vertex~$4$ in $G_4$ to
the vertex~$3$ in $H_4$, which completes the isomorphism 
\[
\phi_2 = \left(
\begin{array}{ccccc}
1 & 2 & 3 & 4 & 5 \\
5 & 1 & 4 & 3 & 2
\end{array}
\right)
\] 
that is in $\iso{G,H}$.  
Finally, both $G_4$ and $H_4$ are 
deleted, and the algorithm 
terminates.~\qed$_{\mbox{\it\small End of Example~\ref{ex}}}$
\end{example}

To prove the correctness of the algorithm, we
argue that:
\begin{description}
\item[(a)] each pair $\pair{\hat{G},\hat{H}}$ constructed in any loop of
  $g$ is a pair of isomorphic graphs---hence, $f$ can legally
  be called in each loop of~$g$; and
  
\item[(b)] the mapping $\phi$ computed by $g$ on input $\pair{G,H}$ is in
  $\iso{G,H}$.
\end{description}

{\bf Proof of~(a):}~~This assertion 
follows immediately from the construction and the
assumption that $G$ and $H$ are isomorphic.

{\bf Proof of~(b):}~~The first call to $f$ yields a valid initial
segment $(x_1, y_1)$ of an isomorphism between $G$ and~$H$, since $f$
is queried about the unmodified graphs $G$ and~$H$.

Let $\phi_i = \{(x_1, y_1), (x_2, y_2), \ldots , (x_i, y_i)\}$ be the
initial segment of $\phi$ that consists of $i$ vertex pairs for
some~$i$, $1 \leq i \leq n$, where $(x_i, y_i)$ is the pair added in the 
$i$~loop of~$g$. 
Let $G_i$ and $H_i$
be the graphs
constructed from $G$ and $H$ when loop $i$ is entered; for example,
$G_1 = G$ and $H_1 = H$.  Fix some $i$ with $1 < i \leq n$.  We show that
the extension $\phi_i$ of $\phi_{i-1}$ (obtained by adding the pair
$(x_i, y_i)$ in the
$i$th loop of~$g$) 
is compatible with $\phi_{i-1}$.  That is, for each
$(x_j,y_j) \in \phi_{i-1}$, it holds that
\[
\{x_i,x_j\} \in E(G) \mbox{ if and only if } \{y_i,y_j\} \in E(H).
\]  

Assume $\{x_i,x_j\} \in E(G)$.  In loop $j < i$, all neighbors
of~$x_j$, including~$x_i$, and all neighbors of $y_j$ were extended to
a clique of size~$n+j-1$.  Note that, in each loop of~$g$, the clique
sizes are increased by one, each clique contains exactly one old
vertex, and any two cliques in $G_i$ (respectively, in~$H_i$) can
overlap only by having their unique old vertex in common.  It follows
that any isomorphism between $G_i$ and $H_i$ must map cliques of size
$n+j-1$ in $G_i$ onto cliques of size $n+j-1$ in~$H_i$.  Since $y_i$
is chosen in loop~$i$ of ~$g$, it follows from our construction that
the clique $C_{n+j-1,x_i}$ in $G_i$ was mapped onto the clique
$C_{n+j-1,y_i}$ in~$H_i$.  Hence, $y_i$ is a neighbor of $y_j$ in~$H$,
i.e., $\{y_i,y_j\} \in E(H)$.  

The converse implication ($\{y_i,y_j\} \in E(H) \,\Longrightarrow\,
\{x_i,x_j\} \in E(G)$) follows by a symmetric argument.

Finally, we estimate the time complexity of the algorithm~$g$.  Since in
each loop of~$g$, 
a pair of old vertices from $V(G) \times V(H)$ is deleted
from the graphs and is added to the isomorphism $\phi \in \iso{G,H}$,
the algorithm terminates after $n$ loops.  Within each loop, $g$ makes
one oracle call to~$f$, updates~$\phi$, and modifies the current
graphs $\hat{G}$ and $\hat{H}$ by deleting certain vertices and by
adding at most $2(n-1)$ cliques of size at most~$2n-1$.  Hence, $g$
runs in cubic time.~\qed

\section{Computing Complete Hamiltonian Cycles from Partial Ones}

Now we turn to the problem of computing complete Hamiltonian cycles in
a graph from partial ones.  Our construction is easier to describe for
multigraphs, i.e., graphs with reflexive and multiple edges allowed.
We may do so, as for Hamiltonian cycles it does not matter whether
simple graphs or multigraphs are used.  We also assume that all graphs
are connected.  

Let us informally describe how our procedure works.  
As in the preceding section, suppose we have a function oracle $f$ that,
given any multigraph $G$ that contains a Hamiltonian cycle, 
returns an edge $e$ that is part of a Hamiltonian cycle $c$ of~$G$.
We want to reduce $G$ by deleting $e$ and
identifying the two vertices incident to $e$, and then want to recursively
apply $f$ to this reduced graph, call it~$\hat{G}$.  However, this approach
would destroy important information about~$e$, namely the ``left'' and the
``right'' context of $e$ in~$G$.  Thus, in the next recursion loop,
the oracle might return an edge contained in a Hamiltonian cycle
$\hat{c}$ of $\hat{G}$ that is not compatible with the previously
chosen edge~$e$, which means
that adding $e$ back to $\hat{G}$
does not necessarily imply that $\hat{c}$ can be extended to a 
Hamiltonian cycle of~$G$.
To preclude cases like this, we require our oracle to return only edges
contained in Hamiltonian cycles that are compatible with the
left-right-context of the edges previously chosen.  
This additional
requirement regarding $f$ makes Theorem~\ref{thm:hc} somewhat weaker than
Theorem~\ref{thm:graphiso}.

First, we define what we mean by a left-right-context of (the edges of)~$G$,
and what we mean by Hamiltonian cycles being compatible (or consistent) with a
left-right-context of~$G$.

\begin{definition}
  Let $G = (V,E)$ be an undirected multigraph with $n$ vertices. 
\begin{itemize}
\item[$\bullet$] A {\em Hamiltonian cycle\/} of $G$ is a sequence $(v_1, v_2,
  \ldots , v_n)$ of pairwise distinct vertices from $V$ such that $\{v_n, v_1\}
  \in E$ and $\{v_i, v_{i+1}\} \in E$ for each~$i$ with $1 \leq i \leq n-1$.
  
\item[$\bullet$] For any set~$S$, let $\mathfrak{P}(S)$ denote the power set of~$S$.
  For any $v \in V$, let $E(v)$ denote the set of edges in $E$
  incident to~$v$.  
  
  A {\em left-right-context of $G$\/} is a function $\pi : V \rightarrow
  \mathfrak{P}(E) \times \mathfrak{P}(E)$ satisfying that, for every $v \in
  \mbox{\rm domain}(\pi)$, there exist sets $L(v)$ and $R(v)$ such that  
\begin{enumerate}
\item $\pi(v) = (L(v), R(v))$,
\item $L(v) \cup R(v) \seq E(v)$, and
\item $L(v) \cap R(v) = \emptyset$.
\end{enumerate}

\item[$\bullet$] We say that a Hamiltonian cycle $c$ of $G$ is {\em consistent with a
    left-right-context $\pi$ of $G$\/} if and only if for every $v \in
  \mbox{\rm domain}(\pi)$, $c$ contains exactly one edge from $L(v)$ and
  exactly one edge from~$R(v)$, where $\pi(v) = (L(v), R(v))$.
\end{itemize}
\end{definition}

We now state our result.

\begin{theorem}
\label{thm:hc}
Let $\hat{G}$ be any multigraph, and let $\pi$ be any
left-right-context of~$\hat{G}$. Suppose there exists a function
oracle $f$ that, given $(\hat{G}, \pi)$, outputs some edge $e \in E(\hat{G})$
such that some Hamiltonian cycle consistent with $\pi$ contains $e$
(provided $\hat{G}$ has a Hamiltonian cycle consistent with~$\pi$).

Then, there is a recursive procedure $g$ that, given any multigraph $G$
that has a Hamiltonian cycle, uses the oracle $f$ to construct a complete
Hamiltonian cycle of $G$ in polynomial time.
\end{theorem}

\noindent
{\bf Proof.}~~Let $G$ be any multigraph with $n$ vertices
that contains a Hamiltonian cycle.  Let $f$ be a function oracle as
in the theorem.  

In the procedure described below, 
whenever we identify two vertices $u$ and~$v$, deleting the edge(s) connecting
$u$ and~$v$, we assume by convention that in the resulting graph the vertex $u
= v$ has two name tags, namely $u$ and~$v$.  This convention simplifies the
description of our construction and does no harm.

We now describe the procedure $g$ on input~$G$:
\begin{description}
\item[Step~0:] Let $G_0 = (V_0, E_0)$ be the given multigraph~$G$, and
  let $\pi_0$
  be the nowhere defined function (on the domain~$V_0$).  
  Set $C$ to the empty set.  Note that $C$ will,
  eventually, contain the complete Hamiltonian cycle of $G$ to be
  constructed.

\item[{\boldmath Step~$i$, $1 \leq i \leq n-1$}:] Let $G_{i-1} =
(V_{i-1}, E_{i-1})$ be the multigraph and let $\pi_{i-1}$ be the
left-right-context of $G_{i-1}$ constructed in the previous step.
Compute the edge $e_i = f(G_{i-1}, \pi_{i-1})$ by querying the oracle,
and add $e_i$ to~$C$.
Let $e_i = \{u_i, v_i\}$.  
Consider the following three cases.
\begin{description}
\item[Case~1:] $e_i \cap \mbox{\rm domain}(\pi_{i-1}) = \emptyset$.

  Cancel $e_i$ from $G_{i-1}$, and identify the vertices $u_i$
  and~$v_i$.
  Call the resulting graph $G_i = (V_i, E_i)$.  Define the left-right-context 
  $\pi_i : V_i \rightarrow \mathfrak{P}(E_i) \times \mathfrak{P}(E_i)$ by
  $\mbox{\rm domain}(\pi_i) = \mbox{\rm domain}(\pi_{i-1}) \cup
  \{u_i\}$ and
\[
\pi_i(v) = \left\{
\begin{array}{ll}
\pi_{i-1}(v) & \mbox{if $v \in \mbox{\rm domain}(\pi_{i-1})$} \\
(L_i(u_i) , R_i(u_i)) & \mbox{if $v = u_i$,}
\end{array}
\right.
\]
where 
\begin{itemize}
\item[$\bullet$] $L_i(u_i) = E_{i-1}(u_i) - \{e_i\}$ and 
\item[$\bullet$] $R_i(u_i) = \{\{u_i, z\}
\condition \{v_i, z\} \in E_{i-1} \ \wedge\ z \neq u_i\}$.
\end{itemize}

\item[Case~2:] $e_i \cap \mbox{\rm domain}(\pi_{i-1}) = \{x\}$ for
  some vertex $x \in V_{i-1}$.  

  By our assumption that $f$ returns only edges consistent with the given
  left-right-context, $e_i$ must belong to
  exactly one of $L_{i-1}(x)$ or
  $R_{i-1}(x)$.  
  Assume $x = v_i$ and
  $e_i \in L_{i-1}(x)$; the other cases---such as the case ``$x = u_i$
  and $e_i \in R_{i-1}(x)$''---can be treated analogously.
  
  Cancel $e_i$ from $G_{i-1}$, and identify the vertices $u_i$ and~$v_i$,
  which equals~$x$.  Call the resulting graph $G_i = (V_i, E_i)$.  
  Define the left-right-context 
  $\pi_i : V_i \rightarrow \mathfrak{P}(E_i) \times \mathfrak{P}(E_i)$ by
  $\mbox{\rm domain}(\pi_i) = \mbox{\rm domain}(\pi_{i-1})$ and
\[
\pi_i(v) = \left\{
\begin{array}{ll}
\pi_{i-1}(v) & \mbox{if $v \neq x$} \\
(L_i(x) , R_i(x)) & \mbox{if $v = x$,}
\end{array}
\right.
\]
where 
\begin{itemize}
\item[$\bullet$] $L_i(x) = \{\{x, z\} \condition \{u_i, z\} \in E_{i-1} \ \wedge\ z \neq
  v_i\}$ and
\item[$\bullet$] $R_i(x) = R_{i-1}(x)$.
\end{itemize}

\item[Case~3:] $e_i \cap \mbox{\rm domain}(\pi_{i-1}) = \{x, y\}$ for
  two vertices $x, y \in V_{i-1}$ with $x \neq y$.  

  It follows that $e_i = \{x, y\}$ in this case.
  By our assumption that $f$ returns only edges consistent with the given
  left-right-context, $e_i$ must belong to exactly one of $L_{i-1}(z)$ or
  $R_{i-1}(z)$, for both $z=x$ and $z=y$.  
  Assume $e_i \in
  L_{i-1}(x) \cap R_{i-1}(y)$; the other cases can be treated
  analogously.
  
  Cancel $e_i$ from $G_{i-1}$, and identify the vertices $x$ and~$y$.
  Call the resulting graph $G_i = (V_i, E_i)$.  Define the left-right-context $\pi_i
  : V_i \rightarrow \mathfrak{P}(E_i) \times \mathfrak{P}(E_i)$ by $\mbox{\rm
    domain}(\pi_i) = \mbox{\rm domain}(\pi_{i-1})$ and
\[
\pi_i(v) = \left\{
\begin{array}{ll}
\pi_{i-1}(v) & \mbox{if $v \neq x = y$} \\
(L_i(y) , R_i(y)) & \mbox{if $v = x = y$,}
\end{array}
\right.
\]
where 
\begin{itemize}
\item[$\bullet$] $L_i(y) = L_{i-1}(y)$ and 
\item[$\bullet$] $R_i(y) = \{\{y, z\} \condition \{x, z\} \in R_{i-1}(x)\}$.
\end{itemize}
\end{description}

\item[{\boldmath Step~$n$}:] Since in each of the $n-1$ previous steps
two vertices have been identified and one edge has been added to~$C$,
the graph $G_{n-1}$ constructed in the previous step contains only one vertex,
say~$z$, having possibly multiple reflexive edges.  Also, $C$ contains
$n-1$ elements, and $\pi_{n-1}$ is either of the form 
\begin{itemize}
\item[$\bullet$] $\pi_{n-1} = (\emptyset, R_{n-1}(z))$ or 
\item[$\bullet$] $\pi_{n-1} = (L_{n-1}(z), \emptyset)$,
\end{itemize}
where any edge in $R_{n-1}(z)$ (respectively, in $L_{n-1}(z)$) can be
used to complete the Hamiltonian cycle constructed so far.  Thus, we may
choose any one edge from $R_{n-1}(z)$ (respectively, from
$L_{n-1}(z)$) and add it to~$C$.
\end{description}
This concludes the description of the procedure~$g$.  Note that $g$ runs
in polynomial time.  To prove the correctness of the algorithm, note
that, for each $i \in \{1, 2, n-2\}$, and for each Hamiltonian cycle $c$
of $G_{i}$ consistent with~$\pi_{i}$, it holds that inserting the edge
$e_i$ into $c$ yields a Hamiltonian cycle of~$G_{i-1}$, 
thus ensuring consistency during of the overall construction.~\qed

\section{Conclusions and Future Work}

In this paper, we studied an important property of NP problems:
how to compute complete solutions from partial
solutions.  
We in particular studied the graph
isomorphism problem and the Hamiltonian cycle problem.  
We showed as
Theorem~\ref{thm:graphiso} that computing even a single pair of vertices
belonging to an isomorphism between two isomorphic graphs is as hard as
computing a complete isomorphism between the graphs.  
Theorem~\ref{thm:graphiso} optimally improves upon a result of G\'{a}l et
al.~\cite{gal-hal-lip-pet:c:partial-solutions}.

We propose to establish
analogous results for NP problems other than the graph isomorphism problem.
For example, G\'{a}l et al.~\cite{gal-hal-lip-pet:c:partial-solutions}
investigated many more hard NP problems, and showed that computing partial
solutions for them is as hard as computing complete solutions.  However, their
results are not known to be optimal, which leaves open the possibility of
improvement.  Relatedly, what impact does the self-reducibility of such
problems have for reducing complete search to partial search?

We obtained as
Theorem~\ref{thm:hc} a similar result about
reducing complete search to partial search for the Hamiltonian cycle problem.  
However, this result appears to
be slightly weaker than Theorem~\ref{thm:graphiso}, 
since in Theorem~\ref{thm:hc} we require a stronger
hypothesis about the function oracle used.  Whether this stronger hypothesis
in fact is necessary remains an open question.  It would be interesting to
know whether, also for the Hamiltonian cycle problem, one can prove a result
as strong as Theorem~\ref{thm:graphiso}.  More precisely, is it possible to
prove the same conclusion as in Theorem~\ref{thm:hc} when we are given a
function oracle that is merely required to return any one edge of a
Hamiltonian cycle of the given graph, without requiring in addition that the
edge returned belong to a Hamiltonian cycle consistent with the edge's
left-right-context?

\bigskip

\noindent
{\bf Acknowledgments.}\quad
We thank Edith and Lane A. Hemaspaandra for introducing us to this
interesting topic and for stimulating discussions and comments.  
We acknowledge interesting discussions about graph theory with 
Haiko M\"uller. 

\bibliography{/sartre2/home2/rothe/BIGBIB/joergbib}

\end{document}

%% file: example.eepic
\setlength{\unitlength}{0.000453333in}
\begingroup\makeatletter\ifx\SetFigFont\undefined%
\gdef\SetFigFont#1#2#3#4#5{%
  \reset@font\fontsize{#1}{#2pt}%
  \fontfamily{#3}\fontseries{#4}\fontshape{#5}%
  \selectfont}%
\fi\endgroup%
{\renewcommand{\dashlinestretch}{30}
\begin{picture}(4280,1995)(0,-10)
\thicklines
\put(1725,1890){\blacken\ellipse{150}{150}}
\put(1725,1890){\ellipse{150}{150}}
\put(1125,1290){\blacken\ellipse{150}{150}}
\put(1125,1290){\ellipse{150}{150}}
\put(1125,690){\blacken\ellipse{150}{150}}
\put(1125,690){\ellipse{150}{150}}
\put(1125,90){\blacken\ellipse{150}{150}}
\put(1125,90){\ellipse{150}{150}}
\put(2925,1290){\blacken\ellipse{150}{150}}
\put(2925,1290){\ellipse{150}{150}}
\put(2925,690){\blacken\ellipse{150}{150}}
\put(2925,690){\ellipse{150}{150}}
\put(3525,690){\blacken\ellipse{150}{150}}
\put(3525,690){\ellipse{150}{150}}
\put(3525,1290){\blacken\ellipse{150}{150}}
\put(3525,1290){\ellipse{150}{150}}
\put(4125,690){\blacken\ellipse{150}{150}}
\put(4125,690){\ellipse{150}{150}}
\path(525,1890)(1125,1290)
\path(525,1890)(1125,1290)
\path(1725,1890)(1125,1290)
\path(1725,1890)(1125,1290)
\path(1125,1290)(1125,690)
\path(1125,1290)(1125,690)
\path(1125,690)(1125,90)
\path(1125,690)(1125,90)
\path(2925,1290)(2925,690)
\path(2925,1290)(2925,690)
\put(525,1890){\blacken\ellipse{150}{150}}
\put(525,1890){\ellipse{150}{150}}
\path(2925,690)(3525,690)
\path(2925,690)(3525,690)
\put(300,1590){\makebox(0,0)[lb]{\smash{{{\SetFigFont{12}{14.4}{\rmdefault}{\mddefault}{\updefault}$1$}}}}}
\path(3525,690)(3525,1290)
\path(3525,690)(3525,1290)
\path(3525,690)(4125,690)
\path(3525,690)(4125,690)
\put(1800,1590){\makebox(0,0)[lb]{\smash{{{\SetFigFont{12}{14.4}{\rmdefault}{\mddefault}{\updefault}$2$}}}}}
\put(1350,1215){\makebox(0,0)[lb]{\smash{{{\SetFigFont{12}{14.4}{\rmdefault}{\mddefault}{\updefault}$3$}}}}}
\put(1350,615){\makebox(0,0)[lb]{\smash{{{\SetFigFont{12}{14.4}{\rmdefault}{\mddefault}{\updefault}$4$}}}}}
\put(1350,15){\makebox(0,0)[lb]{\smash{{{\SetFigFont{12}{14.4}{\rmdefault}{\mddefault}{\updefault}$5$}}}}}
\put(0,915){\makebox(0,0)[lb]{\smash{{{\SetFigFont{12}{14.4}{\rmdefault}{\mddefault}{\updefault}$G$:}}}}}
\put(2250,915){\makebox(0,0)[lb]{\smash{{{\SetFigFont{12}{14.4}{\rmdefault}{\mddefault}{\updefault}$H$:}}}}}
\put(3450,240){\makebox(0,0)[lb]{\smash{{{\SetFigFont{12}{14.4}{\rmdefault}{\mddefault}{\updefault}$4$}}}}}
\put(2850,240){\makebox(0,0)[lb]{\smash{{{\SetFigFont{12}{14.4}{\rmdefault}{\mddefault}{\updefault}$3$}}}}}
\put(4050,240){\makebox(0,0)[lb]{\smash{{{\SetFigFont{12}{14.4}{\rmdefault}{\mddefault}{\updefault}$5$}}}}}
\put(2850,1515){\makebox(0,0)[lb]{\smash{{{\SetFigFont{12}{14.4}{\rmdefault}{\mddefault}{\updefault}$2$}}}}}
\put(3450,1515){\makebox(0,0)[lb]{\smash{{{\SetFigFont{12}{14.4}{\rmdefault}{\mddefault}{\updefault}$1$}}}}}
\end{picture}
}

%% file: badprocedure.eepic
\setlength{\unitlength}{0.000453333in}
\begingroup\makeatletter\ifx\SetFigFont\undefined%
\gdef\SetFigFont#1#2#3#4#5{%
  \reset@font\fontsize{#1}{#2pt}%
  \fontfamily{#3}\fontseries{#4}\fontshape{#5}%
  \selectfont}%
\fi\endgroup%
{\renewcommand{\dashlinestretch}{30}
\begin{picture}(4280,1755)(0,-10)
\thicklines
\put(1725,1650){\blacken\ellipse{150}{150}}
\put(1725,1650){\ellipse{150}{150}}
\put(1125,1050){\blacken\ellipse{150}{150}}
\put(1125,1050){\ellipse{150}{150}}
\put(1125,450){\blacken\ellipse{150}{150}}
\put(1125,450){\ellipse{150}{150}}
\put(2925,450){\blacken\ellipse{150}{150}}
\put(2925,450){\ellipse{150}{150}}
\put(3525,450){\blacken\ellipse{150}{150}}
\put(3525,450){\ellipse{150}{150}}
\put(3525,1050){\blacken\ellipse{150}{150}}
\put(3525,1050){\ellipse{150}{150}}
\put(4125,450){\blacken\ellipse{150}{150}}
\put(4125,450){\ellipse{150}{150}}
\path(525,1650)(1125,1050)
\path(525,1650)(1125,1050)
\path(1725,1650)(1125,1050)
\path(1725,1650)(1125,1050)
\path(1125,1050)(1125,450)
\path(1125,1050)(1125,450)
\path(2925,450)(3525,450)
\path(2925,450)(3525,450)
\put(525,1650){\blacken\ellipse{150}{150}}
\put(525,1650){\ellipse{150}{150}}
\path(3525,450)(3525,1050)
\path(3525,450)(3525,1050)
\put(3450,1275){\makebox(0,0)[lb]{\smash{{{\SetFigFont{12}{14.4}{\rmdefault}{\mddefault}{\updefault}$1$}}}}}
\path(3525,450)(4125,450)
\path(3525,450)(4125,450)
\put(1800,1350){\makebox(0,0)[lb]{\smash{{{\SetFigFont{12}{14.4}{\rmdefault}{\mddefault}{\updefault}$2$}}}}}
\put(1350,975){\makebox(0,0)[lb]{\smash{{{\SetFigFont{12}{14.4}{\rmdefault}{\mddefault}{\updefault}$3$}}}}}
\put(1350,375){\makebox(0,0)[lb]{\smash{{{\SetFigFont{12}{14.4}{\rmdefault}{\mddefault}{\updefault}$4$}}}}}
\put(0,675){\makebox(0,0)[lb]{\smash{{{\SetFigFont{12}{14.4}{\rmdefault}{\mddefault}{\updefault}$\widehat{G}$:}}}}}
\put(2250,675){\makebox(0,0)[lb]{\smash{{{\SetFigFont{12}{14.4}{\rmdefault}{\mddefault}{\updefault}$\widehat{H}$:}}}}}
\put(300,1350){\makebox(0,0)[lb]{\smash{{{\SetFigFont{12}{14.4}{\rmdefault}{\mddefault}{\updefault}$1$}}}}}
\put(2850,0){\makebox(0,0)[lb]{\smash{{{\SetFigFont{12}{14.4}{\rmdefault}{\mddefault}{\updefault}$3$}}}}}
\put(3450,0){\makebox(0,0)[lb]{\smash{{{\SetFigFont{12}{14.4}{\rmdefault}{\mddefault}{\updefault}$4$}}}}}
\put(4050,0){\makebox(0,0)[lb]{\smash{{{\SetFigFont{12}{14.4}{\rmdefault}{\mddefault}{\updefault}$5$}}}}}
\end{picture}
}

%% file: goodprocedure.eepic
\setlength{\unitlength}{0.000453333in}
\begingroup\makeatletter\ifx\SetFigFont\undefined%
\gdef\SetFigFont#1#2#3#4#5{%
  \reset@font\fontsize{#1}{#2pt}%
  \fontfamily{#3}\fontseries{#4}\fontshape{#5}%
  \selectfont}%
\fi\endgroup%
{\renewcommand{\dashlinestretch}{30}
\begin{picture}(4880,2295)(0,-10)
\thicklines
\put(1725,2190){\blacken\ellipse{150}{150}}
\put(1725,2190){\ellipse{150}{150}}
\put(1125,1590){\blacken\ellipse{150}{150}}
\put(1125,1590){\ellipse{150}{150}}
\put(1125,990){\blacken\ellipse{150}{150}}
\put(1125,990){\ellipse{150}{150}}
\put(3525,990){\blacken\ellipse{150}{150}}
\put(3525,990){\ellipse{150}{150}}
\put(4125,990){\blacken\ellipse{150}{150}}
\put(4125,990){\ellipse{150}{150}}
\put(675,690){\ellipse{150}{150}}
\put(1575,690){\ellipse{150}{150}}
\put(825,90){\ellipse{150}{150}}
\put(1425,90){\ellipse{150}{150}}
\put(4725,990){\blacken\ellipse{150}{150}}
\put(4725,990){\ellipse{150}{150}}
\put(4125,1590){\blacken\ellipse{150}{150}}
\put(4125,1590){\ellipse{150}{150}}
\put(3075,690){\ellipse{150}{150}}
\put(3975,690){\ellipse{150}{150}}
\put(3225,90){\ellipse{150}{150}}
\put(3825,90){\ellipse{150}{150}}
\path(525,2190)(1125,1590)
\path(525,2190)(1125,1590)
\path(1725,2190)(1125,1590)
\path(1725,2190)(1125,1590)
\path(1125,1590)(1125,990)
\path(1125,1590)(1125,990)
\path(3525,990)(4125,990)
\path(3525,990)(4125,990)
\path(1125,990)(675,690)
\path(675,690)(825,90)
\path(825,90)(1425,90)
\path(1425,90)(1575,690)
\path(1575,690)(1125,990)
\path(1125,990)(825,90)
\put(525,2190){\blacken\ellipse{150}{150}}
\put(525,2190){\ellipse{150}{150}}
\path(825,90)(1575,690)
\put(2675,1215){\makebox(0,0)[lb]{\smash{{{\SetFigFont{12}{14.4}{\rmdefault}{\mddefault}{\updefault}$H_1$:}}}}}
\path(1575,690)(675,690)
\path(675,690)(1425,90)
\path(1425,90)(1125,990)
\path(4125,1590)(4125,990)
\path(4125,1590)(4125,990)
\path(4125,990)(4725,990)
\path(4125,990)(4725,990)
\path(3525,990)(3075,690)
\path(3075,690)(3225,90)
\path(3225,90)(3825,90)
\path(3825,90)(3975,690)
\path(3975,690)(3525,990)
\path(3525,990)(3225,90)
\path(3225,90)(3975,690)
\path(3975,690)(3075,690)
\path(3075,690)(3825,90)
\path(3825,90)(3525,990)
\put(1800,1890){\makebox(0,0)[lb]{\smash{{{\SetFigFont{12}{14.4}{\rmdefault}{\mddefault}{\updefault}$2$}}}}}
\put(1350,1515){\makebox(0,0)[lb]{\smash{{{\SetFigFont{12}{14.4}{\rmdefault}{\mddefault}{\updefault}$3$}}}}}
\put(1350,915){\makebox(0,0)[lb]{\smash{{{\SetFigFont{12}{14.4}{\rmdefault}{\mddefault}{\updefault}$4$}}}}}
\put(-100,1215){\makebox(0,0)[lb]{\smash{{{\SetFigFont{12}{14.4}{\rmdefault}{\mddefault}{\updefault}$G_1$:}}}}}
\put(300,1890){\makebox(0,0)[lb]{\smash{{{\SetFigFont{12}{14.4}{\rmdefault}{\mddefault}{\updefault}$1$}}}}}
\put(4650,540){\makebox(0,0)[lb]{\smash{{{\SetFigFont{12}{14.4}{\rmdefault}{\mddefault}{\updefault}$5$}}}}}
\put(4050,1815){\makebox(0,0)[lb]{\smash{{{\SetFigFont{12}{14.4}{\rmdefault}{\mddefault}{\updefault}$1$}}}}}
\put(4200,1140){\makebox(0,0)[lb]{\smash{{{\SetFigFont{12}{14.4}{\rmdefault}{\mddefault}{\updefault}$4$}}}}}
\put(3450,1140){\makebox(0,0)[lb]{\smash{{{\SetFigFont{12}{14.4}{\rmdefault}{\mddefault}{\updefault}$3$}}}}}
\end{picture}
}

%% file: goodprocedure2.eepic
\setlength{\unitlength}{0.00043333in}
\begingroup\makeatletter\ifx\SetFigFont\undefined%
\gdef\SetFigFont#1#2#3#4#5{%
  \reset@font\fontsize{#1}{#2pt}%
  \fontfamily{#3}\fontseries{#4}\fontshape{#5}%
  \selectfont}%
\fi\endgroup%
{\renewcommand{\dashlinestretch}{30}
\begin{picture}(5190,2370)(0,-10)
\thicklines
\put(1125,1590){\blacken\ellipse{150}{150}}
\put(1125,1590){\ellipse{150}{150}}
\put(1125,990){\blacken\ellipse{150}{150}}
\put(1125,990){\ellipse{150}{150}}
\put(3525,990){\blacken\ellipse{150}{150}}
\put(3525,990){\ellipse{150}{150}}
\put(675,690){\ellipse{150}{150}}
\put(1575,690){\ellipse{150}{150}}
\put(825,90){\ellipse{150}{150}}
\put(1425,90){\ellipse{150}{150}}
\put(3075,690){\ellipse{150}{150}}
\put(3975,690){\ellipse{150}{150}}
\put(3225,90){\ellipse{150}{150}}
\put(3825,90){\ellipse{150}{150}}
\put(525,2190){\ellipse{150}{150}}
\put(450,1815){\ellipse{150}{150}}
\put(900,2265){\ellipse{150}{150}}
\put(750,1515){\ellipse{150}{150}}
\put(1200,1965){\ellipse{150}{150}}
\put(4425,990){\blacken\ellipse{150}{150}}
\put(4425,990){\ellipse{150}{150}}
\put(4425,1590){\blacken\ellipse{150}{150}}
\put(4425,1590){\ellipse{150}{150}}
\put(5025,390){\ellipse{150}{150}}
\put(4350,615){\ellipse{150}{150}}
\put(4800,1065){\ellipse{150}{150}}
\put(4650,315){\ellipse{150}{150}}
\put(5100,765){\ellipse{150}{150}}
\path(1725,2190)(1125,1590)
\path(1725,2190)(1125,1590)
\path(1125,1590)(1125,990)
\path(1125,1590)(1125,990)
\path(1125,990)(675,690)
\path(675,690)(825,90)
\path(825,90)(1425,90)
\path(1425,90)(1575,690)
\path(1575,690)(1125,990)
\path(1125,990)(825,90)
\path(825,90)(1575,690)
\path(1575,690)(675,690)
\path(675,690)(1425,90)
\path(1425,90)(1125,990)
\path(3525,990)(3075,690)
\path(3075,690)(3225,90)
\path(3225,90)(3825,90)
\path(3825,90)(3975,690)
\path(3975,690)(3525,990)
\path(3525,990)(3225,90)
\path(3225,90)(3975,690)
\put(1725,2190){\blacken\ellipse{150}{150}}
\put(1725,2190){\ellipse{150}{150}}
\path(3975,690)(3075,690)
\put(2550,1215){\makebox(0,0)[lb]{\smash{{{\SetFigFont{12}{14.4}{\rmdefault}{\mddefault}{\updefault}$H_2$:}}}}}
\path(3075,690)(3825,90)
\path(3825,90)(3525,990)
\path(525,2190)(450,1815)
\path(450,1815)(750,1515)
\path(750,1515)(1125,1590)
\path(1125,1590)(1200,1965)
\path(1200,1965)(900,2265)
\path(900,2265)(525,2190)
\path(525,2190)(750,1515)
\path(750,1515)(1200,1965)
\path(1200,1965)(450,1815)
\path(450,1815)(900,2265)
\path(900,2265)(750,1515)
\path(525,2190)(1200,1965)
\path(450,1815)(1125,1590)
\path(1125,1590)(900,2265)
\path(525,2190)(1125,1590)
\path(4425,1665)(4425,1065)
\path(4425,1665)(4425,1065)
\path(3525,990)(4425,990)
\path(4425,990)(4350,615)
\path(4350,615)(4650,315)
\path(4650,315)(5025,390)
\path(5025,390)(5100,765)
\path(5100,765)(4800,1065)
\path(4800,1065)(4425,990)
\path(4425,990)(4650,315)
\path(4650,315)(5100,765)
\path(5100,765)(4425,990)
\path(4350,615)(5025,390)
\path(5025,390)(4800,1065)
\path(4800,1065)(4350,615)
\path(4350,615)(5100,765)
\path(4800,1065)(4650,315)
\path(4425,990)(5025,390)
\put(1800,1890){\makebox(0,0)[lb]{\smash{{{\SetFigFont{12}{14.4}{\rmdefault}{\mddefault}{\updefault}$2$}}}}}
\put(1350,1515){\makebox(0,0)[lb]{\smash{{{\SetFigFont{12}{14.4}{\rmdefault}{\mddefault}{\updefault}$3$}}}}}
\put(1350,915){\makebox(0,0)[lb]{\smash{{{\SetFigFont{12}{14.4}{\rmdefault}{\mddefault}{\updefault}$4$}}}}}
\put(-150,1215){\makebox(0,0)[lb]{\smash{{{\SetFigFont{12}{14.4}{\rmdefault}{\mddefault}{\updefault}$G_2$:}}}}}
\put(3450,1140){\makebox(0,0)[lb]{\smash{{{\SetFigFont{12}{14.4}{\rmdefault}{\mddefault}{\updefault}$3$}}}}}
\put(4125,1140){\makebox(0,0)[lb]{\smash{{{\SetFigFont{12}{14.4}{\rmdefault}{\mddefault}{\updefault}$4$}}}}}
\put(4125,1815){\makebox(0,0)[lb]{\smash{{{\SetFigFont{12}{14.4}{\rmdefault}{\mddefault}{\updefault}$1$}}}}}
\end{picture}
}

%% file: goodprocedure3.eepic
\setlength{\unitlength}{0.00043333in}
\begingroup\makeatletter\ifx\SetFigFont\undefined%
\gdef\SetFigFont#1#2#3#4#5{%
  \reset@font\fontsize{#1}{#2pt}%
  \fontfamily{#3}\fontseries{#4}\fontshape{#5}%
  \selectfont}%
\fi\endgroup%
{\renewcommand{\dashlinestretch}{30}
\begin{picture}(5415,2370)(0,-10)
\thicklines
\put(1350,990){\blacken\ellipse{150}{150}}
\put(1350,990){\ellipse{150}{150}}
\put(3750,990){\blacken\ellipse{150}{150}}
\put(3750,990){\ellipse{150}{150}}
\put(900,690){\ellipse{150}{150}}
\put(1800,690){\ellipse{150}{150}}
\put(1050,90){\ellipse{150}{150}}
\put(1650,90){\ellipse{150}{150}}
\put(3300,690){\ellipse{150}{150}}
\put(4200,690){\ellipse{150}{150}}
\put(3450,90){\ellipse{150}{150}}
\put(4050,90){\ellipse{150}{150}}
\put(750,2190){\ellipse{150}{150}}
\put(675,1815){\ellipse{150}{150}}
\put(1125,2265){\ellipse{150}{150}}
\put(975,1515){\ellipse{150}{150}}
\put(1425,1965){\ellipse{150}{150}}
\put(4650,990){\blacken\ellipse{150}{150}}
\put(4650,990){\ellipse{150}{150}}
\put(5250,390){\ellipse{150}{150}}
\put(4575,615){\ellipse{150}{150}}
\put(5025,1065){\ellipse{150}{150}}
\put(4875,315){\ellipse{150}{150}}
\put(5325,765){\ellipse{150}{150}}
\put(1650,1890){\ellipse{150}{150}}
\put(1650,1290){\ellipse{150}{150}}
\put(2250,1815){\ellipse{150}{150}}
\put(2250,1365){\ellipse{150}{150}}
\put(1950,1965){\ellipse{150}{150}}
\put(1950,1215){\ellipse{150}{150}}
\put(4350,1290){\ellipse{150}{150}}
\put(4950,1290){\ellipse{150}{150}}
\put(4275,1590){\ellipse{150}{150}}
\put(5025,1590){\ellipse{150}{150}}
\put(4425,1890){\ellipse{150}{150}}
\put(4875,1890){\ellipse{150}{150}}
\path(1350,1590)(1350,990)
\path(1350,1590)(1350,990)
\path(1350,990)(900,690)
\path(900,690)(1050,90)
\path(1050,90)(1650,90)
\path(1650,90)(1800,690)
\path(1800,690)(1350,990)
\path(1350,990)(1050,90)
\path(1050,90)(1800,690)
\path(1800,690)(900,690)
\path(900,690)(1650,90)
\path(1650,90)(1350,990)
\path(3750,990)(3300,690)
\path(3300,690)(3450,90)
\path(3450,90)(4050,90)
\path(4050,90)(4200,690)
\path(4200,690)(3750,990)
\path(3750,990)(3450,90)
\path(3450,90)(4200,690)
\path(4200,690)(3300,690)
\path(3300,690)(4050,90)
\path(4050,90)(3750,990)
\path(750,2190)(675,1815)
\path(675,1815)(975,1515)
\path(975,1515)(1350,1590)
\path(1350,1590)(1425,1965)
\path(1425,1965)(1125,2265)
\path(1125,2265)(750,2190)
\path(750,2190)(975,1515)
\path(975,1515)(1425,1965)
\path(1425,1965)(675,1815)
\path(675,1815)(1125,2265)
\path(1125,2265)(975,1515)
\path(750,2190)(1425,1965)
\path(675,1815)(1350,1590)
\put(1350,1590){\blacken\ellipse{150}{150}}
\put(1350,1590){\ellipse{150}{150}}
\path(1350,1590)(1125,2265)
\put(4025,1815){\makebox(0,0)[lb]{\smash{{{\SetFigFont{12}{14.4}{\rmdefault}{\mddefault}{\updefault}$v$}}}}}
\path(750,2190)(1350,1590)
\path(3750,990)(4650,990)
\path(4650,990)(4575,615)
\path(4575,615)(4875,315)
\path(4875,315)(5250,390)
\path(5250,390)(5325,765)
\path(5325,765)(5025,1065)
\path(5025,1065)(4650,990)
\path(4650,990)(4875,315)
\path(4875,315)(5325,765)
\path(5325,765)(4650,990)
\path(4575,615)(5250,390)
\path(5250,390)(5025,1065)
\path(5025,1065)(4575,615)
\path(4575,615)(5325,765)
\path(5025,1065)(4875,315)
\path(4650,990)(5250,390)
\path(1350,1590)(1650,1890)
\path(1650,1890)(1950,1965)
\path(1950,1965)(2250,1815)
\path(2250,1815)(2250,1365)
\path(2250,1365)(1950,1215)
\path(1950,1215)(1650,1290)
\path(1650,1290)(1350,1590)
\path(1350,1590)(1950,1965)
\path(1350,1590)(2250,1815)
\path(1350,1590)(2250,1365)
\path(1350,1590)(1950,1215)
\path(1650,1890)(2250,1815)
\path(1650,1890)(2250,1365)
\path(1650,1890)(1950,1215)
\path(1650,1890)(1650,1290)
\path(1950,1965)(2250,1365)
\path(1950,1965)(1950,1215)
\path(1950,1965)(1650,1290)
\path(2250,1815)(1950,1215)
\path(2250,1815)(1650,1290)
\path(2250,1365)(1650,1290)
\path(4650,990)(4350,1290)
\path(4350,1290)(4275,1590)
\path(4275,1590)(4425,1890)
\path(4425,1890)(4875,1890)
\path(4875,1890)(5025,1590)
\path(5025,1590)(4950,1290)
\path(4950,1290)(4650,990)
\path(4650,990)(4275,1590)
\path(4650,990)(4425,1890)
\path(4650,990)(4875,1890)
\path(4650,990)(5025,1590)
\path(4350,1290)(4425,1890)
\path(4350,1290)(4875,1890)
\path(4350,1290)(5025,1590)
\path(4350,1290)(4950,1290)
\path(4275,1590)(4875,1890)
\path(4275,1590)(5100,1590)
\path(4275,1590)(4950,1290)
\path(4425,1890)(5025,1590)
\path(4425,1890)(4950,1290)
\path(4875,1890)(4950,1290)
\put(1575,915){\makebox(0,0)[lb]{\smash{{{\SetFigFont{12}{14.4}{\rmdefault}{\mddefault}{\updefault}$4$}}}}}
\put(3675,1140){\makebox(0,0)[lb]{\smash{{{\SetFigFont{12}{14.4}{\rmdefault}{\mddefault}{\updefault}$3$}}}}}
\put(1125,1215){\makebox(0,0)[lb]{\smash{{{\SetFigFont{12}{14.4}{\rmdefault}{\mddefault}{\updefault}$3$}}}}}
\put(0,1215){\makebox(0,0)[lb]{\smash{{{\SetFigFont{12}{14.4}{\rmdefault}{\mddefault}{\updefault}$G_3$:}}}}}
\put(2625,1215){\makebox(0,0)[lb]{\smash{{{\SetFigFont{12}{14.4}{\rmdefault}{\mddefault}{\updefault}$H_3$:}}}}}
\put(4350,690){\makebox(0,0)[lb]{\smash{{{\SetFigFont{12}{14.4}{\rmdefault}{\mddefault}{\updefault}$4$}}}}}
\put(2475,1740){\makebox(0,0)[lb]{\smash{{{\SetFigFont{12}{14.4}{\rmdefault}{\mddefault}{\updefault}$u$}}}}}
\end{picture}
}

%% file: goodprocedure4.eepic
\setlength{\unitlength}{0.00043333in}
\begingroup\makeatletter\ifx\SetFigFont\undefined%
\gdef\SetFigFont#1#2#3#4#5{%
  \reset@font\fontsize{#1}{#2pt}%
  \fontfamily{#3}\fontseries{#4}\fontshape{#5}%
  \selectfont}%
\fi\endgroup%
{\renewcommand{\dashlinestretch}{30}
\begin{picture}(4440,2295)(0,-10)
\thicklines
\put(3750,990){\blacken\ellipse{150}{150}}
\put(3750,990){\ellipse{150}{150}}
\put(900,690){\ellipse{150}{150}}
\put(1800,690){\ellipse{150}{150}}
\put(1050,90){\ellipse{150}{150}}
\put(1650,90){\ellipse{150}{150}}
\put(3300,690){\ellipse{150}{150}}
\put(4200,690){\ellipse{150}{150}}
\put(3450,90){\ellipse{150}{150}}
\put(4050,90){\ellipse{150}{150}}
\put(1350,2190){\ellipse{150}{150}}
\put(750,1590){\ellipse{150}{150}}
\put(1950,1590){\ellipse{150}{150}}
\put(900,2040){\ellipse{150}{150}}
\put(900,1140){\ellipse{150}{150}}
\put(1800,1140){\ellipse{150}{150}}
\put(1800,2040){\ellipse{150}{150}}
\put(3750,2190){\ellipse{150}{150}}
\put(3150,1590){\ellipse{150}{150}}
\put(4350,1590){\ellipse{150}{150}}
\put(3300,2040){\ellipse{150}{150}}
\put(4200,2040){\ellipse{150}{150}}
\put(3300,1140){\ellipse{150}{150}}
\put(4200,1140){\ellipse{150}{150}}
\path(1350,990)(900,690)
\path(900,690)(1050,90)
\path(1050,90)(1650,90)
\path(1650,90)(1800,690)
\path(1800,690)(1350,990)
\path(1350,990)(1050,90)
\path(1050,90)(1800,690)
\path(1800,690)(900,690)
\path(900,690)(1650,90)
\path(1650,90)(1350,990)
\path(3750,990)(3300,690)
\path(3300,690)(3450,90)
\path(3450,90)(4050,90)
\path(4050,90)(4200,690)
\path(4200,690)(3750,990)
\path(3750,990)(3450,90)
\path(3450,90)(4200,690)
\path(4200,690)(3300,690)
\path(3300,690)(4050,90)
\path(4050,90)(3750,990)
\path(1350,990)(900,1140)
\path(900,1140)(750,1590)
\path(750,1590)(900,2040)
\path(900,2040)(1350,2190)
\path(1350,2190)(1800,2040)
\path(1800,2040)(1950,1590)
\path(1950,1590)(1800,1140)
\put(1350,990){\blacken\ellipse{150}{150}}
\put(1350,990){\ellipse{150}{150}}
\path(1800,1140)(1350,990)
\put(0,990){\makebox(0,0)[lb]{\smash{{{\SetFigFont{12}{14.4}{\rmdefault}{\mddefault}{\updefault}$G_4$:}}}}}
\path(1350,990)(750,1590)
\path(900,2040)(1350,990)
\path(1350,990)(1350,2190)
\path(1800,2040)(1350,990)
\path(1350,990)(1950,1590)
\path(900,1140)(900,2040)
\path(1350,2190)(900,1140)
\path(900,1140)(1800,2040)
\path(1950,1590)(900,1140)
\path(900,1140)(1800,1140)
\path(750,1590)(1350,2190)
\path(1800,2040)(750,1590)
\path(750,1590)(1950,1590)
\path(1800,1140)(750,1590)
\path(900,2040)(1800,2040)
\path(1950,1590)(900,2040)
\path(900,2040)(1800,1140)
\path(1350,2190)(1950,1590)
\path(1350,2190)(1800,1140)
\path(1800,2040)(1800,1140)
\path(3750,990)(3300,1140)
\path(3300,1140)(3150,1590)
\path(3150,1590)(3300,2040)
\path(3300,2040)(3750,2190)
\path(3750,2190)(4200,2040)
\path(4200,2040)(4350,1590)
\path(4350,1590)(4200,1140)
\path(4200,1140)(3750,990)
\path(3750,990)(3150,1590)
\path(3300,2040)(3750,990)
\path(3750,990)(3750,2190)
\path(4200,2040)(3750,990)
\path(3750,990)(4350,1590)
\path(3300,1140)(3300,2040)
\path(3750,2190)(3300,1140)
\path(3300,1140)(4200,2040)
\path(4350,1590)(3300,1140)
\path(3300,1140)(4200,1140)
\path(3150,1590)(3750,2190)
\path(4200,2040)(3150,1590)
\path(3150,1590)(4350,1590)
\path(4200,1140)(3150,1590)
\path(3300,2040)(4200,2040)
\path(4350,1590)(3300,2040)
\path(3300,2040)(4200,1140)
\path(3750,2190)(4350,1590)
\path(4200,1140)(3750,2190)
\path(4200,2040)(4200,1140)
\put(2400,990){\makebox(0,0)[lb]{\smash{{{\SetFigFont{12}{14.4}{\rmdefault}{\mddefault}{\updefault}$H_4$:}}}}}
\end{picture}
}